# Design Study of APS-U-Type Hybrid-MBA Lattice for Mid-Energy DLSR

Yu Zhao[1,2,3], Yi Jiao[1,2] *, Sheng Wang[1,2,3] *

1. Institute of High Energy Physics (IHEP), Chinese Academy of Sciences (CAS)，Beijing 100049, China
2. University of Chinese Academy of Sciences (UCAS), Beijing 100049, China
3. Spallation Neutron Source Science Center (SNSSC), Dongguan 523803, China

* Corresponding authors. E-mail addresses: jiaoyi@ihep.ac.cn (Yi Jiao), wangs@ihep.ac.cn (Sheng Wang)

**Abstract**

In recent years, a new generation of storage ring-based light sources, known as diffraction-limited storage rings (DLSRs), whose emittance approaches the diffraction limit for the range of X-ray wavelengths of interest to the scientific community, has garnered significant attention worldwide. Researchers have begun to design and build DLSRs. Among various DLSR proposals, the hybrid multibend achromat (H-MBA) lattice enables sextupole strengths to be maintained at a reasonable level when minimizing the emittance; hence, it has been adopted in many DLSR designs. Based on the H-7BA lattice, the design of the Advanced Photon Source Upgrade Project (APS-U) can effectively reduce emittance by replacing six quadrupoles with anti-bends. Herein, we discuss the feasibility of designing an APS-U-type H-MBA lattice for the Southern Advanced Photon Source, a mid-energy DLSR light source with ultralow emittance that has been proposed to be built adjacent to the China Spallation Neutron Source. Both linear and nonlinear dynamics are optimized to obtain a detailed design of this type of lattice. The emittance is minimized, while a sufficiently large dynamic aperture (DA) and momentum acceptance (MA) are maintained. A design comprising 36 APS-U type H-7BAs, with an energy of 3 GeV and a circumference of 972 m, is achieved. The horizontal natural emittance is 20 pm·rad, with a horizontal DA of 5.8 mm, a vertical DA of 4.5 mm, and an MA of 4%, as well as a long longitudinal damping time of 120 ms. Subsequently, a few modifications are performed based on the APS-U-type lattice to reduce the longitudinal damping time from 120 to 44 ms while maintaining other performance parameters at the same level.

## 1 Introduction

To fulfill the continuous demand for higher brightness and better coherence in X-rays, storage-ring-based light sources have been developed for three generations, in addition to the development of accelerator physics and technology [1]. A new generation of storage-ring-based light sources, known as diffraction-limited storage rings (DLSRs) [2], has been proposed to surpass the brightness and coherence limits of existing third-generation light sources using a multibend achromat (MBA) lattice in the design to reduce the emittance such that the limits approach the diffraction limit for multi-kilo electron-volt (keV) photons.

Except the MBA [3] lattices used in MAX IV [4] and Sirius [5] designs, the H-MBA [6] lattice and modified H-MBA lattices (e.g., [7-9]), which are innovative variations of the MBA lattice, as will be introduced briefly in next section, have been widely adopted in the designs of DLSRs, particularly in high-energy DLSR designs.

However, many options are available for the lattice of low- and mid-energy storage rings (e.g., the 7BA lattice for PEP-X [10] and H-6BA lattice for the Diamond-II [9]). The "APS-U -type H-MBA" lattice [7], a concept that combines anti-bends (ABs) into a H-MBA, has been rarely used in mid-energy DLSR designs, particularly designs involving a long circumference. To investigate the feasibility of designing a mid-energy DLSR using the APS-U-type H-MBA lattice, a lattice of this type was designed for the Southern Advanced Photon Source (SAPS) in this study. The SAPS is a mid-energy DLSR light source near the China Spallation Neutron Source in Guangdong province, China.

The paper is organized as follows: The linear optics design and nonlinear optimization for the SAPS are introduced in Sections 3 and 4, respectively. It is discovered that in addition to the decrease in emittance, the APS-U-type lattice design for the SAPS may become impractical owing to the extremely long damping time when the emittance approaches the diffraction limit for multi-keV photons. In Section 5, some measures for managing the abovementioned issue are introduced, and conclusions are presented in Section 6.

## 2 Advent of APS-U-type H-MBA lattice

The horizontal natural emittance of electron beam in a storage ring can be written as [11]

$$\varepsilon_0 = C_q F(type) \frac{\gamma^2}{J_x N_b^3}, \qquad (1)$$

where $C_q = 3.83 \times 10^{-13}$ m, $N_b$ is the number of bending magnets, $\gamma$ is the Lorentz factor, $J_x$ is the horizontal damping partition number, and $F(type)$ is a dimensionless quantity that depends on the lattice type. For example, $F$ is $1/12\sqrt{15}$ for a theoretical minimum emittance (TME) unit cell [11]. As shown in Eq. (1), the emittance can be reduced by increasing the number of bending magnets; therefore, the MBA lattice was selected as the basic layout for the DLSR design.

In fact, an MBA lattice comprising five or more bending magnets per achromat has been investigated as early as the 1990s [3]. Nevertheless, it remained challenging to materialize the MBA lattice, until the development of magnets with ultra-high gradients [12] and improved vacuum in an ultra-small aperture [13], as well as the adoption of these techniques in the design of MAX IV [4] with an emittance of 326 pm·rad. One MBA comprises M-2 (where M is the number of bending magnets per achromat) modified TME unit cells [14] and two matching cells. In each modified TME unit cell, high-gradient horizontally focusing quadrupoles and a dipole combined with a horizontally defocusing gradient (BD) are adopted to reduce the horizontal natural emittance and the unit cell length. Two matching cells containing at least two quadrupoles are used to match the dispersion function to be zero at the long straight section. When the emittance of such an MBA lattice design is reduced continuously, the dispersion function decreases and the natural chromaticity increases; hence, stronger and bulkier sextupoles are required for chromaticity correction, which may limit the compactness of such a design [15].

Hence, a hybrid-MBA (H-MBA) comprising M-4 modified TME unit cells and two DBA-like cells, was proposed and first used in the design of the European Synchrotron Radiation Facility upgrade project [6]. Using a H-7BA as an example, chromatic sextupoles are used in the

DBA-like cells, where dispersion bumps are created between the first and second dipoles (as well as between the sixth and seventh dipoles in symmetry) to maintain the magnetic strengths of the sextupoles to a reasonable level. These dipoles are longitudinal gradient dipoles (LGBs) [16], and each LGB is composed of five pieces, which are considered to have different bending angles. They are generally designed such that for the piece closer to the dispersion bump, the corresponding bending angle is smaller. In addition, the phase advance between each pair of sextupoles is designed to be at or close to $n\pi$ ($n$ is an odd integer) [17] to eliminate most of the nonlinear effects caused by the sextupoles. Typically, the phase advance is $3\pi$ in the horizontal direction and $\pi$ in the vertical direction.

Based on the H-MBA concept, many modified H-MBA lattices have been developed and adopted in the designs of DLSR light sources, such as APS-U [7], High Energy Photon Source (HEPS) [18], Diamond-II [9] and Pohang Accelerator Laboratory - four-generation storage ring (PAL-4GSR) [19], which are summarized in Table 1. Some combined function magnets and super-bends were used in the modified H-MBA lattice designs to further reduce the emittance. In the design of PAL-4GSR [19], super-bends with magnetic fields up to 2 T (similar to super-bends with magnetic fields up to 3.2 T in the design of Sirius [5]) were used to reduce the emittance and served as bending magnet sources to provide high-quality X-rays. Meanwhile, in the APS-U design, anti-bends [20] (which are bends with reverse bending angles that allow the independent control of beta and dispersion functions) were used to replace six quadrupoles of the H-7BA lattice, thereby reducing the emittance from 67 to 41 pm·rad [7].

**Table 1** Properties of some DLSR light sources adopting modified hybrid-MBAs
(symbol "+" implies that some combined function magnets are combined into hybrid-MBA)

| Parameters | APS-U | HEPS | PAL-4GSR | Diamond-II |
|---|---|---|---|---|
| Energy (GeV) | 6 | 6 | 3 | 3.5 |
| Circumference (m) | 1104 | 1360.4 | 570 | 560.6 |
| Emittance (pm·rad) | 41.7 | 34.8 | 89.4 | 157 |
| Lattice structure | H-7BA + 6ABs | H-7BA + 4ABs + LGB | H-7BA + 6ABs + super-bends | H-6BA + 4ABs |

## 3  Linear optics design

The current lattice design for DLSRs is more complex than that for third-generation light sources, considering the diversity, quantity and complexity of the magnets involved. Moreover, engineering limits of the magnets and the balance among the objective functions must be considered. Consequently, the design process often relies on identification from a large parameter space, in which the use of efficient algorithms and heavy computation are expected. Therefore, DLSR designs are typically optimized using stochastic optimization algorithms, e.g., the multi-objective genetic algorithm [21-25], multi-objective particle swarm optimization (MOPSO) algorithm [26-29], or machine-learning-enhanced stochastic optimization algorithms [30-32]. In this study, we used MOPSO to optimize the linear dynamics and performed the optimization procedure proposed in Ref. [33].

In the linear optimization of the lattice, the sextupole strengths can be used as an indicator of the nonlinear performance. However, previous studies [e.g., 21, 34] demonstrated that, if two objectives (emittance and chromatic sextupole strengths or emittance and natural chromaticity) are used in the optimization, then the solutions obtained with minimized sextupole strengths or natural chromaticity do not definitely have large dynamic apertures (DAs). In Ref. [33], it is shown that if three objectives (emittance, natural chromaticity, and chromatic sextupole strengths) are used, then the balance between natural chromaticity and chromatic sextupole strength will aid in obtaining a lattice with a large DA when the emittance is minimized. Therefore, in this study, for the linear lattice design of the SAPS (based on Ref. [33]), we used the following three objective functions: the horizontal natural emittance $\varepsilon_0$, sum of the absolute strengths of the sextupoles ($|K|_{sum}$), and sum of the absolute values of natural chromaticities ($|\zeta_x| + |\zeta_y|$).

The APS-U-type H-7BA lattice of the SAPS, as mentioned above, combines six anti-bends (four in the middle of the dispersion bump, and the other two on each side of the fourth dipole, marked as Q4/5/8 in Fig. 1(a)) into the H-7BA lattice. In the linear optimization of such a lattice, all tunable parameters (exceeding 30), except the lengths of quadrupoles, sextupoles, anti-bends (which are adjusted to ensure that the magnetic parameters are within the engineering limits), and the long straight sections, are scanned. Some constraints for the magnets based on the limits of the HEPS [35, 36], as well as the limits for the drift spaces are listed in Table 2.

It is noteworthy that the fractional part of the transverse tune must be sufficiently far from the integer and half-integer resonances. The chromaticities were corrected to [+5,+5] with three families of sextupoles (SD1, SD2, and SF) to avoid collective instabilities [37]. Similar to the settings in Ref. [33], SD1 and SF were used to correct 2/3 of the chromaticities, whereas SD2 and SF were used to correct the remainder of the chromaticities. The horizontal and vertical phase advances between each pair of sextupoles were matched to be $3\pi$ to eliminate most of the nonlinear effects caused by the sextupoles.

**Table 2** Constraints for optimization

| Element | Parameters | Values or Limits | Units |
|---|---|---|---|
| Quadrupole | pole face field | < 1 | T |
| | bore radius | 12.5 | mm |
| | gradient | < 80 | T/m |
| Sextupole | pole face field | < 0.6 | T |
| | bore radius | 12.5 | mm |
| | gradient | < 7500 | T/m$^2$ |
| Combined function dipole | bending radius | > 20 | m |
| | gradient | < 48 | T/m |
| | pole face gap | 38 | mm |
| Longitudinal gradient bend | magnetic field | < 1 | T |
| Drift | distance between elements | ≥ 0.1 | m |
| | long straight section | 5 | m |
| | drift of the third or fifth dipole | > 0.35 | m |

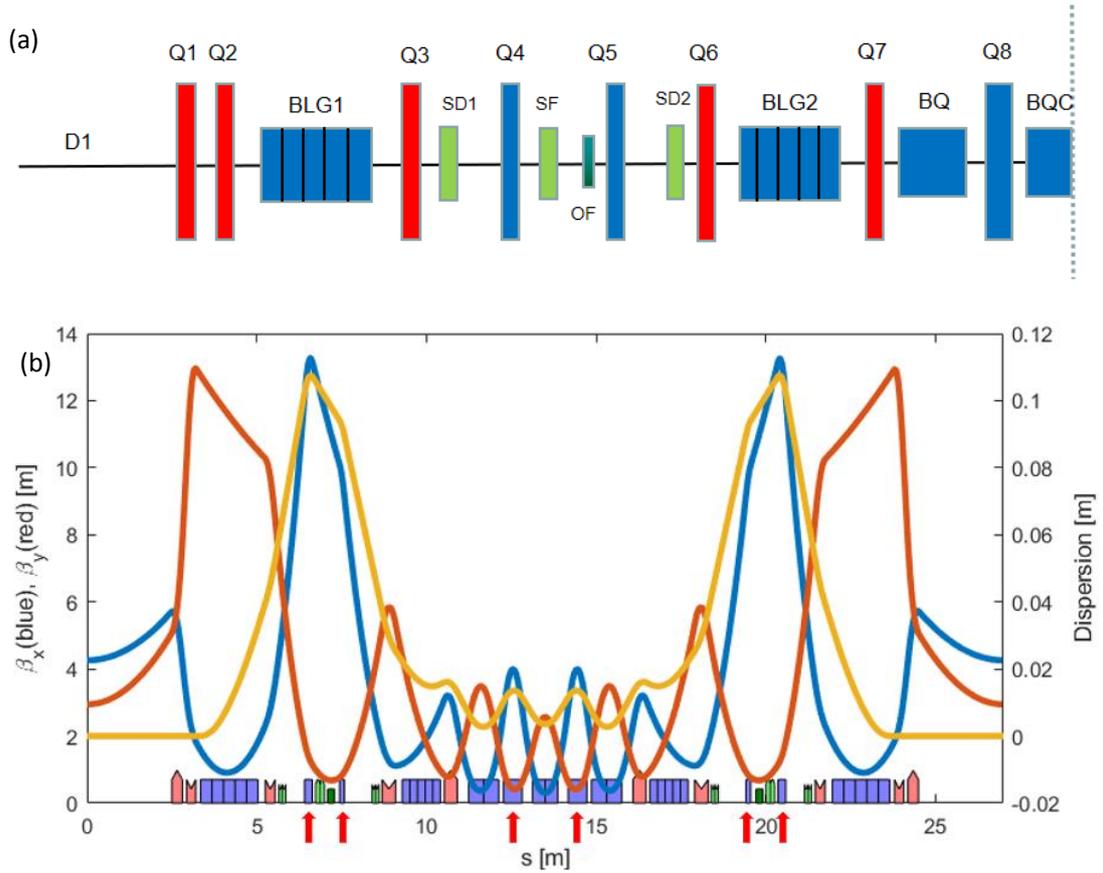

**Fig. 1** Layout (a) and optical functions (b) of APS-U-type H-7BA lattice designed for SAPS (replaced quadrupoles are marked with red arrows). Blue, red, green, and dark green blocks represent dipoles, quadrupoles, sextupoles and octupoles, respectively

Based on the constraints mentioned above, the optimization was performed with a population of 3000 and iterated for 500 generations. The distribution of the objective function values of the last generation is shown in Fig. 2. Although it is shown in Fig. 2 that the minimum emittance is 15 pm·rad, a brief review reveals that the momentum acceptance (MA) for this emittance is extremely low ($DA_x$ ~4 mm and $DA_y$ ~2.5 mm; MA ~0.6%). If the emittance requirement is loosened slightly, then more practical designs can be achieved. For example, a lattice, referred to as Design A, with an emittance of 20 pm·rad, was selected from the optimization solutions owing to its better nonlinear performance ($DA_x$ ~4 mm and $DA_y$ ~4.2 mm; MA ~2.4%) compared with those of other candidates with similar emittances. The optical functions of the selected lattice are shown in Fig. 1 (b). The dipole field profiles of the two pairs of LGBs in Design A, which might not be optimal, are shown in Fig. 3. Nevertheless, from a physics perspective, the profiles may be reasonable owing to the trend of a stronger bending field at the part farther from the dispersion bump. The optimization results above show that a greater horizontal or vertical tune enables a lattice with a lower emittance to be obtained. Moreover, a lattice with minimized chromatic sextupole strengths might not exhibit a favorable nonlinear acceptance, which is consistent with the findings reported in Ref. [22]. In addition, the anti-bend angles added in the H-MBA lattice can generally reduce the emittance; however, a larger anti-bend angle might not imply a lower emittance. In Design A, the total bending angle of the ring increased from 360° to 392.32° by adopting anti-bends. Owing to the relatively high dispersion in the DBA-like cells where chromatic sextupoles are located, the sextupoles gradients and lengths were controlled within a reasonable range (i.e., the lengths of the SD and SF were 0.2 and 0.25 m, respectively, with the gradients being less than 2200 T/m$^2$). More parameters of the SAPS design are listed in Table 3. In fact, the cell length of Design A was set similar to that of the APS-U to investigate the feasibility of designing a lattice composed of an APS-U-type H-MBA with a circumference of approximately 1000 m. Considering that the frequency of the RF cavity was 166.7 MHz and the harmonic number was 540, the cell length was set to 27 m, which is excessively long for a 3 GeV DLSR; nevertheless, this setting minimized the emittance and provided a good basis for lattice modification.

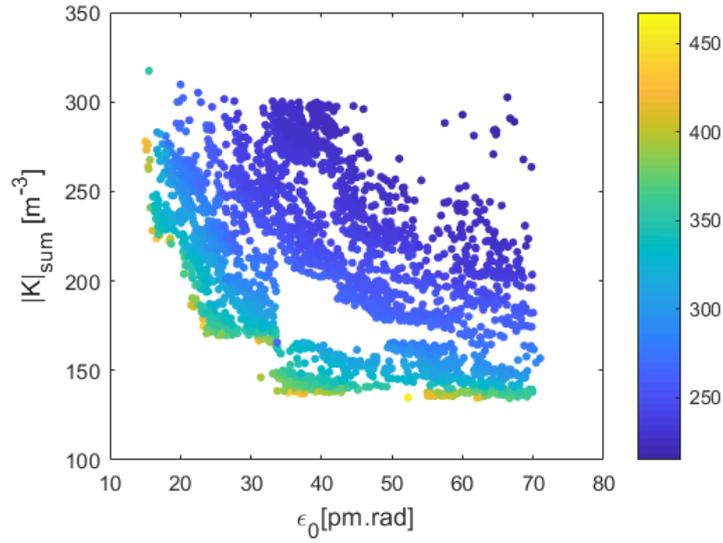

**Fig. 2** Distribution of objective function values of last generation (color bar represents sum of absolute values of natural chromaticities)

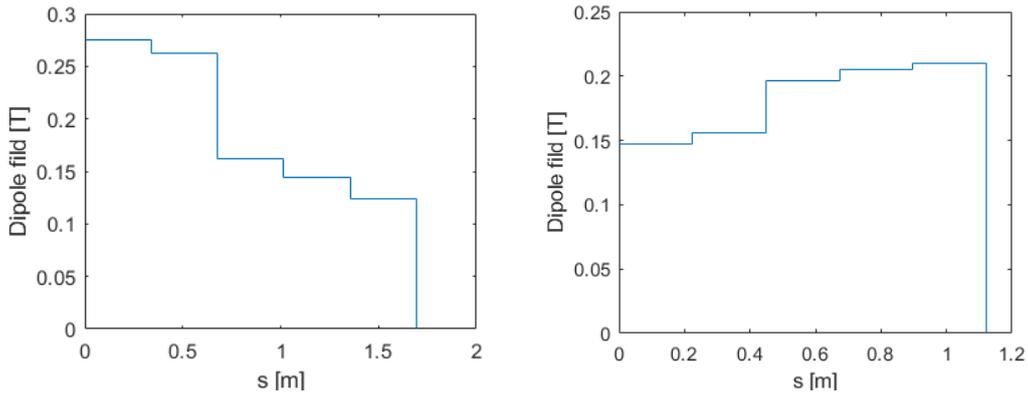

**Fig. 3** Dipole field profiles of LGBs for Design A. Left: dipole field profile of first LGB; right: dipole field profile of second LGB

Table 3  Main parameters of designed lattice

| Parameters | Values (Design A) | Values (Design B) | Units |
|---|---|---|---|
| **Beam energy** | 3 | 3 | GeV |
| **Lattice structure** | APS-U-type H-7BA | Modified APS-U-type H-7BA | |

| Natural emittance | 20 | 23 | pm·rad |
|---|---|---|---|
| Circumference | 972 | 1080 | m |
| Natural energy spread | 6.89e-4 | 9.63e-4 | |
| Length of straight section | 5 | 5 | m |
| RF frequency | 166.7 | 166.7 | MHz |
| RF voltage | 1.2 | 1.2 | MV |
| Corrected chromaticity (H/V) | +5/+5 | +5/+5 | |
| Momentum compaction factor | 3.53e-5 | 1.37e-5 | |
| Harmonic number | 540 | 600 | |
| Natural bunch length | 3.4 | 3.1 | mm |
| Betatron tune (H/V) | 85.11/75.19 | 81.23/64.18 | |
| Radiation energy loss per turn | 0.21 | 0.48 | MeV/turn |
| Damping partition [$x/y/z$] | 2.23/1/0.77 | 1.55/1/1.45 | |
| Damping time [$x/y/z$] | 41.7/93/120.7 | 28.8/44/30.7 | ms |

## 4 Nonlinear optimization

The DA and MA are the two most important objectives for DLSR nonlinear dynamics optimization. A small DA may result in difficult beam injections, whereas a low MA implies a short Touschek lifetime. The linear design obtained in this study exhibited a sufficient DA and a low MA of 2%. However, this provides a good basis for nonlinear dynamics optimization. It had been suggested [29] that using more variables for nonlinear optimization can yield higher values of DA and MA. Therefore, in the nonlinear optimization, except three families of sextupoles that were required to be regrouped to increase the free knobs, two octupoles were introduced adjacent to the horizontally focusing sextupoles (SF), as shown in Fig. 1(a), to further optimize the DA and MA. Among the three families of sextupoles, SD1 and SF were used to correct the chromaticities to [+5, +5]. The strength of SD2 and its position (the distance between SD2 and Q6 in Fig. 1(a)), combined with the strengths of the octupoles were scanned to identify a better MA and DA.

Additionally, the nonlinear dynamics optimization was performed using MOPSO, with a population size of 100 and evolution exceeding 20 generations. A small population and a few iterations can be used since the variables were few. The constraints used in the nonlinear optimization were the same as those in the linear optimization. The distribution of the objective function values with evolution generations is shown in Fig. 4, and the result marked with a black box was selected as the final optimization result. Fig. 5 shows the tune shifts with momentum deviation after the optimization, indicating that the MA increased to 4%. Compared with an MA of only 2.4% from the linear optimization, this is a significant improvement. To calculate the local MA, 6D particle tracking over 1024 turns was performed for one cell in Design A, assuming that the frequency of the RF cavity was 166.7 MHz and its voltage was 1.2 MV. The result is shown in Fig. 6. The minimum MA exceeded 2.85%.

Fig. 7 presents the on-momentum DA at the center of the long straight section and the corresponding frequency map, obtained from 4D particle tracking over 1024 turns where a bare lattice was considered. The horizontal and vertical DAs were approximately 5.8 and 4.5 mm, respectively. Compared with the values of the linear design, both the horizontal and vertical DAs increased. Typically, as the momentum deviation increases, the DA will decrease accordingly. Fig. 8 shows the variations in the DA when the momentum deviation changed from -4% to 4%. For example, when the momentum deviation is 4%, the horizontal and vertical DAs reduce to only approximately 2.2 mm. For such a small DA, an on-axis "swap-out" injection [38] or a longitudinal injection [e.g. 39-43] can be used for the beam injection.

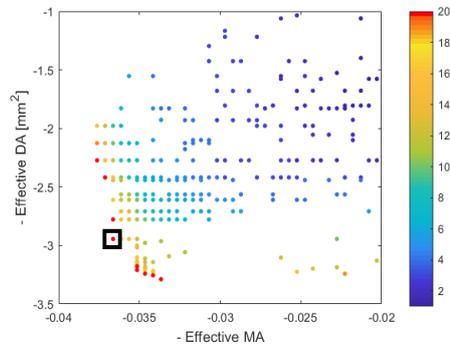

Fig. 4  Distribution of objective function values with evolution generations, where different generations are represented by different colors, and selected result is marked with black box. Within the effective DA and MA, motion must remain stable, and tune footprint is bounded by integer and half-integer resonances closest to working point.

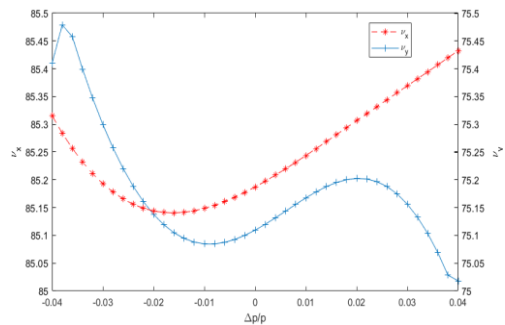

**Fig. 5**  Tune shifts with momentum deviation for Design A

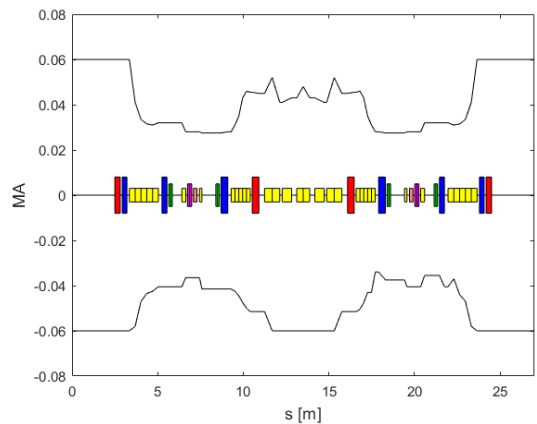

**Fig. 6**  Local MA along one cell of Design A, tracked over 1024 turns with 6D tracking

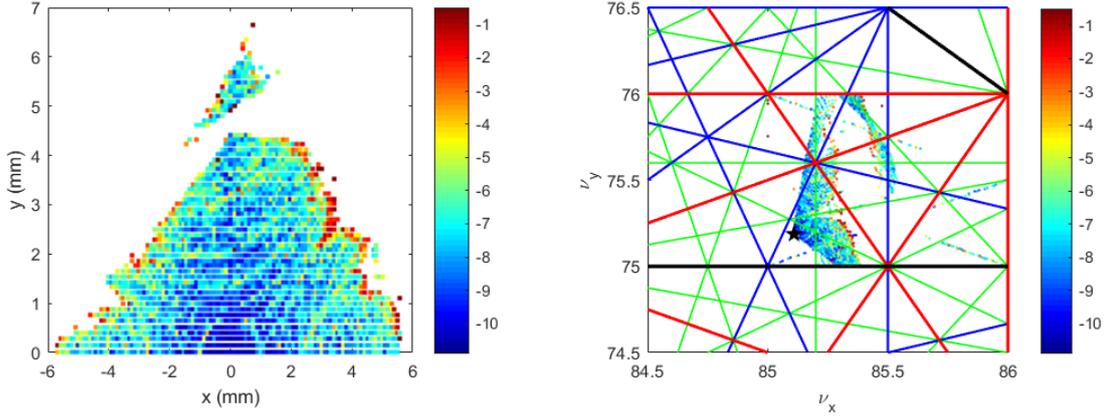

**Fig. 7** Dynamic aperture and frequency map analysis obtained after tracking over 1024 turns for Design A (color bar represents stability of particles; blue and red imply more regular and chaotic motions, respectively)

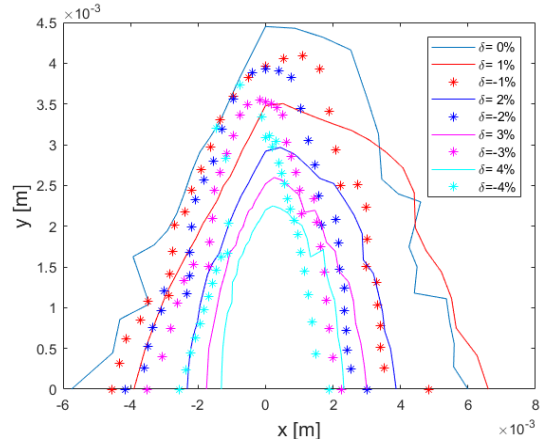

**Fig. 8** Off-momentum DAs obtained after tracking over 1024 turns for Design A

## 5 Damping time

As shown in Eq. (1), the horizontal natural emittance $\varepsilon_0$ scales approximately as $E^2/N_b^3$. For a 7BA lattice with fixed magnet dimensions, $\varepsilon_0 \propto E^2/(7 \times N)^3$, where $N$ represents the number of 7BAs. Moreover, the damping time $\tau$ scales as $N^3/E^3$. The APS-U design is composed of 40 APS-U-type H-7BAs with an emittance of 41.7 pm·rad at 6 GeV, and a longitudinal damping time of 20.5 ms [7]. By scaling down the energy and number of 7BAs to 3 GeV and 36, respectively, i.e., the same numbers as those of the SAPS design, the emittance and longitudinal damping time will be 14.3 pm·rad and 119.6 ms, respectively. The results obtained from the optimization above are not significantly different from the estimated values.

However, the longitudinal damping time of 120 ms may result in emittance deterioration owing to severe collective effects, particularly the intrabeam scattering effect [37]. The damping time can be reduced using damping wigglers. However, a longer damping time necessitates the use of more damping wigglers; hence, more long straight sections will be occupied. Accordingly, the number of long straight sections for insertion devices will be reduced, which is contrary to our expectation. A method to reduce the damping time without using damping wigglers is to decrease the number of 7BAs or use shorter rings; accordingly, the emittance will increase. For instance, in the case of a 540 m ring composed of only 20 APS-U-type H-7BAs, the longitudinal damping time will be 20.5 ms via simple scaling. However, the emittance will be 116.5 pm·rad, which is approximately five times higher than that of Design A.

Another method to reduce the damping time is to increase the energy loss per turn [36], which can be expressed as

$$U_0[\text{keV}] = 88.5 \frac{E^4[\text{GeV}]}{\rho[\text{m}]}, \tag{2}$$

where $E$ is the energy, $\rho = L/\theta$, $L$ is the dipole length, and $\theta$ is the bending angle. The energy loss per turn can be increased by reducing the radii of the dipoles or adding more anti-bends. Based on the current design (Design A) for the SAPS, we replaced the third and fifth dipoles in every achromat with novel combined function dipoles (such as those in Elettra-II [44]) and changed the fourth dipole to a LGB whose central slice exhibited a magnetic field of up to 2 T. Each of the novel combined function dipoles comprised a high-field (up to 1.05 T without a transverse gradient) dipole in the middle and moderate-field (0.31 T with a transverse gradient up to 20 T/m) dipoles on both sides. In addition, two BDs and two high-gradient horizontally focusing anti-bends were adjacent to the LGB. The modifications based on the APS-U-type lattice are illustrated in Fig. 9. Accordingly, the circumference of the new design (referred to as Design B) increased slightly to 1080 m. After a similar linear lattice optimization, the emittance of Design B was minimized to 23 pm·rad, whereas the longitudinal damping time was successfully reduced to 44 ms. More parameters for Design B are listed in Table 3. A similar nonlinear optimization was performed to ensure that the DA and MA remained in a favorable range. Finally, the horizontal DA was 5 mm, vertical DA was 3.5 mm, and MA was 4%, as shown in Fig. 10.

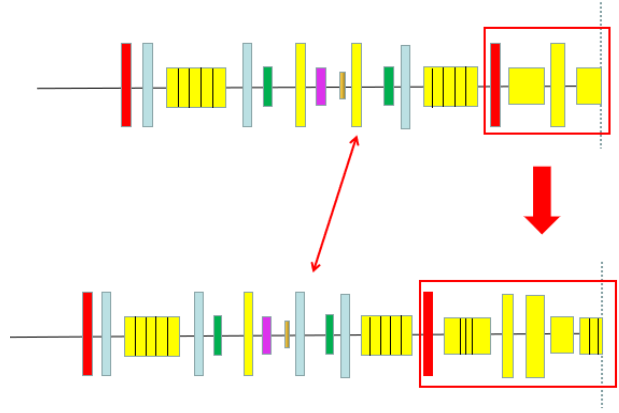

**Fig. 9** Half layouts of Designs A and B. Lattice was symmetric with respect to dotted line. Yellow, red, blue, green, purple, and brown blocks represent dipoles, horizontally focusing and defocusing quadrupoles, horizontally focusing and defocusing sextupoles, and octupoles, respectively. Replaced cells marked with red box, and replaced dipole marked with red arrow

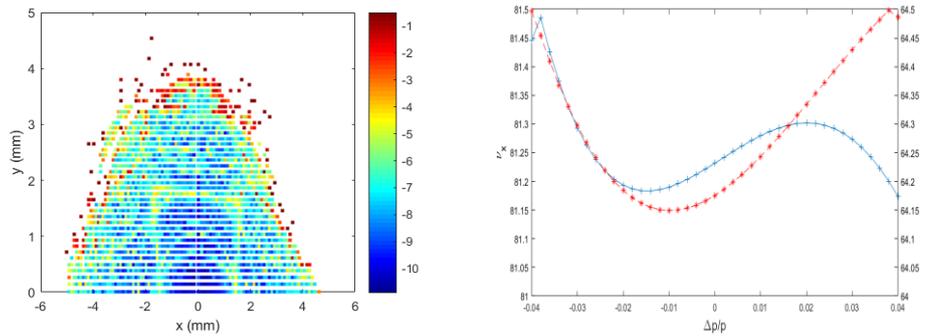

**Fig. 10** Dynamic aperture obtained after tracking over 1024 turns and tune shifts with momentum deviation for Design B (left: DA, right: tune shifts with momentum deviation)

## 6  Conclusions

In this study, a detailed design of a mid-energy ring for the SAPS with an APS-U-type H-MBA lattice was achieved. An ultra-low emittance of approximately 20 pm·rad was achieved with a circumference of approximately 1000 m, and the DA and MA were sufficient for on-axis beam injection. However, the excessively long longitudinal damping time caused the emittance to increase rapidly owing to collective effects, rendering the design unappealing. However, we demonstrated that the longitudinal damping time can be controlled to a reasonable level using a novel combination of function dipoles, LGBs, and anti-bends. We expect the presented results to serve as a useful reference for future SAPS designs and other mid-energy light source designs involving the APS-U-type H-MBA lattice.

**Acknowledge**

This study was supported by the National Natural Science Foundation of China (No. 11922512), the Youth Innovation Promotion Association of Chinese Academy of Sciences (No. Y201904) and the Guangdong Basic and Applied Basic Research Foundation - Guangdong Dongguan Joint Foundation (No. 2019B1515120069).